\documentclass[twocolumn,preprintnumbers,amsmath,amssymb]{revtex4-2}
\usepackage{graphicx}% Include figure files
\usepackage{dcolumn}% Align table columns on decimal point
\usepackage{bm}% bold math
\usepackage{appendix}
\usepackage[dvipsnames]{xcolor}

\newcommand{\ket}[1]{|#1\rangle}

\usepackage{verbatim}
\begin{document}

%\preprint{BCS}

%\preprint{APS/123-QED}

\title{Co-existence of charge density wave and anti-ferromagnetic coupling in the spin-chain compound Ba$_6$Cr$_2$S$_{10}$}
\author{Jianhua Zhu$^{1,2}$}\altaffiliation{These authors contributed equally to this work.}\email{ucapjhz@ucl.ac.uk}
\author{Jianfeng Zhang$^{3,4}$}\altaffiliation{These authors contributed equally to this work.}%\email{zjf@iphys.ac.cn}
\author{Yilin Zhang$^1$}
\author{Devashibhai Adroja$^{5,6}$}
\author{Huancheng Yang$^4$}%\email{hcyang@ruc.edu.cn}
\author{Xiancheng Wang$^3$}
\author{Changqing Jin$^3$}
\author{Ji Chen$^{1,7,8}$}\email{ji.chen@pku.edu.cn}
\author{Wei Wu$^{2}$}\email{wei.wu@ucl.ac.uk}

\affiliation{$^1$School of Physics, Peking University, Chengfu Road 209, Haidian, Beijing 100871, China}
\affiliation{$^2$UCL Department of Physics and Astronomy and London Centre for Nanotechnology, University College London, Gower Street, London WC1E 6BT}
\affiliation{$^3$Beijing National Laboratory for Condensed Matter Physics, Institute of Physics, Chinese Academy of Sciences, Beijing 100190, China}
\affiliation{$^4$Key Laboratory of Quantum State Construction and Manipulation (Ministry of Education), Renmin University of China, Beijing, 100872, China}
\affiliation{$^5$ISIS Neutron and Muon Facility, STFC, Rutherford Appleton Laboratory, Chilton, Didcot, Oxford, OX11 0QX, United Kingdom} \affiliation{$^6$Highly Correlated Matter Research Group, Physics Department, University of Johannesburg, Auckland Park 2006, South Africa}
\affiliation{$^7$Interdisciplinary Institute of Light-Element Quantum Materials 
\\
and Research Center for Light-Element Advanced Materials, Peking University, Beijing 100871, P. R. China}
\affiliation{$^8$Frontiers Science Center for Nano-Optoelectronics, Peking University, Beijing 100871, P. R. China}

\date{\today}%
\begin{abstract}
%The spin chain compound Ba$_6$Cr$_2$S$_{10}$ was synthesised experimentally under high temperature and pressure recently, which showed anti-ferromagnetic magnetic properties along the spin chain. However, the origin of anti-ferromagnetic coupling and electronic structure are still yet to clarify completely. 

Here we have performed detailed first principles calculations for the electronic structure and magnetic properties of Ba$_6$Cr$_2$S$_{10}$ to study the origin of the anti-ferromagnetic exchange interaction between spins on Cr ions for the spin-chain compound Ba$_6$Cr$_2$S$_{10}$ synthesised recently. Most importantly, we have found the co-existence of a charge density wave phase along one line and an anti-ferromagnetic spin chain along another. The dimerization of sulfur atoms loosely bonded with Ba atoms drives the system into an insulating state owing to the formation of charge density wave. Meanwhile, the small size of the effective Hubbard $U$ parameter ($\sim 0.5$ eV) due to electrostatic screening mainly accounts for the anti-ferromagnetic ground state. This co-existence equips us with a platform to tune the charge and spin degrees of freedom independently. Moreover, there exists a next-nearest-neighbouring anti-ferromagnetic interaction along the chain, which could bring forward spin frustration and hence quantum spin liquid. 
\end{abstract}

%\pacs{}
% PACS, the Physics and Astronomy Classification Scheme.
%\keywords{Suggested keywords}%Use showkeys class option if keyword
%display desired

\maketitle

\section{Introduction} 
%Quasi-one-dimensional (quasi-1D) magnetic systems can give rise to many interesting phenomena in condensed matter physics such as quantum spin liquid \cite{tsvelikbook}. 
In one dimension (1D), quantum fluctuations become more important as compared with the thermal ones at sufficiently low temperature and can be manifested in quantum spin liquid\cite{tsvelikbook}, quantum tunnelling \cite{abel2021}, fractional edge states \cite{mishra2021}, and topological order \cite{haldane1983, chen2010}. Quasi-1D spin chains can provide a fascinating physical platform not only to the study of low-dimensional physics but also long-range quantum entanglement \cite{cgw2011,wu2021}. One-dimensional Majorana zero-mode bound states share the same central charge of $\frac12$ with that in the 1D Ising model with transverse magnetic field, as suggested by the conformal field theory \cite{calabrese2004,dalum2020}. One of the most interesting temperature regions is probably where the three-dimensional ordering is not yet established but the temperature is lower than the Curie intercept $\theta_p$ originating from the exchange interaction within the spin chains, provided intra-chain coupling is much stronger than inter-chain, which could lead to quantum spin liquid \cite{tsvelikbook}. Moreover, magnetic toroidal moments - an interesting dipole moment that violates both the spatial and time inversion symmetries can be found in ordered spin systems, for example in 1D \cite{ederer2007, zhang2022}, orbital loop current \cite{varma2006,fsh2006}, and wheel-shaped molecules such as Dy$_6$ \cite{ungur2012,wu2018}. Quasi-1D anti-ferromagnetic (AFM) dimerized spin chains can be used to realize polar toroidal moments in solid state \cite{ederer2007,zhang2021, zhang2022}. 1D spin chains can also find applications in quantum information processing; quantum communications through 1D AFM spin chains have been proposed to realise quantum state transfer between the chain ends \cite{bose2007}. Molecular single-chain magnets can have many potential applications in spintronics and magnetism \cite{bogani2008}.

Recently the compounds consisting of spin chains, either organic or inorganic, have attracted much attention for magnetism and spintronics. Organic spin-chain compounds, in which spin chains weakly interact with each other, can form in phthalocyanine-based molecular chains such as cobalt phthalocyanines \cite{sandrine2007, wu2007, wang2010, wu2011, serri2014,wu2013,wu2013_lipc,wu2013_crpc,wu2014_cocupc}. More broadly speaking, organic single chain magnets, which can explore chemical synthesis and different building blocks for magnetic spin units, have comparable functionalities to single molecule magnets, thus pointing to many potential applications in spintronics and quantum computing \cite{bogani2008}. There are many molecular material candidates for single chain magnets, such as [Mn(TPP)O$_2$PPhH]$\cdot$H$_2$O (TPP = meso-tetraphenylporphyrin), apart from the phthalocyanine-based spin chains \cite{bernot2008}. For inorganic compounds, the AMX$_3$ and B$_3$MM'O$_6$ (A = Cs/Rb, B = alkaline-earth metal, M/M' = transition metals, X = halides) compounds form large families of spin chain compounds, which have exhibited interesting low-dimensional magnetic behaviour, as observed in elastic or inelastic neutron scattering experiments \cite{yelon1975,agrestini2008, toth2016}. CsCoBr$_3$ and Ca$_3$Co$_2$O$_6$ consist of strongly coupled Ising chains, which are vertical to the triangular lattice plane \cite{mao2002}. The spins in the chain in CsCoBr$_3$ are anti-ferromagnetically coupled, while those in Ca$_3$Co$_3$O$_6$ are ferromagnetically arranged \cite{mao2002}. Especially the B$_3$MM'O$_6$ compound family has complex magnetic structures, multiple exchange interactions between spins on the transition-metal ions. In some compounds, $4d$/$5d$ orbitals will be involved, which implies that the spin-orbit interaction will play an important role in magnetic properties such as spin anisotropy \cite{toth2016}. For B$_3$MM'O$_6$ compounds, Sr$_3$ZnRhO$_6$\cite{znrh}, Sr3CuIrO$_6$\cite{cuir}, Sr$_3$CuRhO$_6$\cite{curh}, Ca$_3$CoRhO$_6$\cite{corh}, and Sr$_3$NiIrO$_6$\cite{toth2016} have been extensively studied, showing interesting magnetic properties such as magnetisation jumps and anisotropic exchange interactions. Furthermore, when there are two different types of magnetic atoms in the chain ($3d/4d$ or $3d/5d$), the system exhibits hidden magnetic transition well below the long range ordering and the origin of this is still an open debate \cite{lcs2014}.

\begin{figure}[htbp]
%\begin{tabular}{cc}
%\includegraphics[width=9cm,height=5cm]{fig_5.eps}\\
\includegraphics[width=8cm, height=4.5cm, trim={0cm 0cm 0.0cm 0.0cm},clip]{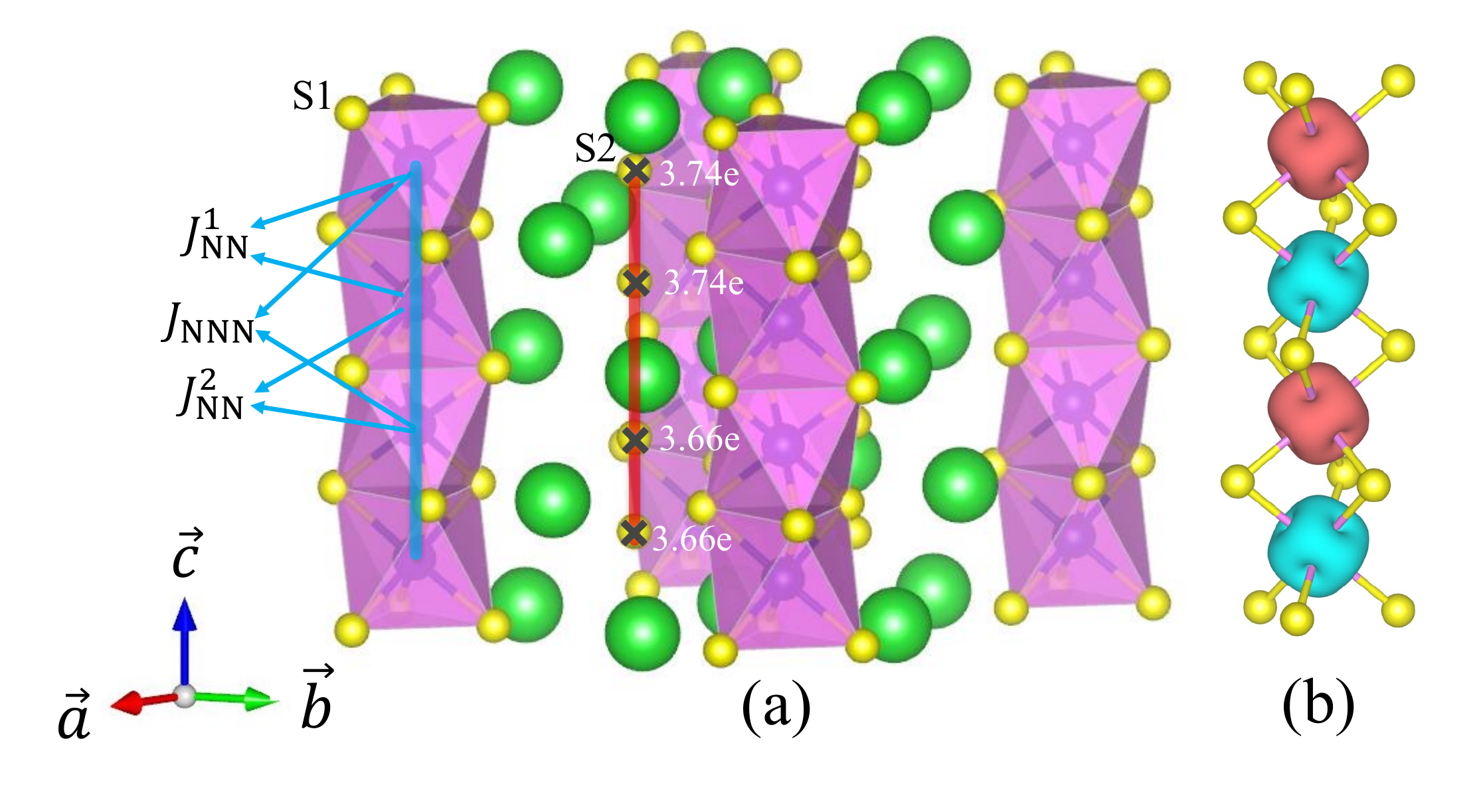}\\
%\textbf{(a)}\\
%\includegraphics[width=8cm,height=7cm]{}\\
%\includegraphics[width=18cm, height=12.35cm, trim={0.cm 0cm 0cm 0cm},clip]{fig_1b.pdf}\\
%\textbf{(b)}\\
%\end{tabular}
\caption{(Colour online.) (a) The crystal structure of BCS is shown. Ba is in green, Cr is in blue, and S is in yellow. Cr atoms, bonded by S atoms, form octahedron-face-sharing chains along the $c$-axis (illustrated by a blue line). S1 and S2 label the sulfur atoms for those forming ligands with Cr and those loosely bonded with Ba (lablled by $\times$; the red line illustrates the S chain), respectively. We have also provided the atomic charges for the S atoms loosely bonded with Ba. The nearest-neighboring exchange interactions $J_{\text{NN}}^1$ and $J_{\text{NN}}^2$, and the next-nearest-neighboring exchange interaction $J_{\text{NNN}}$ have been indicated as well. (b) Spin densities on Cr suggest the orbital ordering on Cr along the chain. Due to the zig-zag ligand, the Cr orbitals will alternate along the $c-$axis.  }\label{fig:1} 
\end{figure}

Hf$_5$Sn$_3$Cu-anti type ternary compounds A$_3$BX$_5$ (A = alkali earth metal, B = transition metal and X = chalcogen) are a typical spin chain compound family that also are candidates for one-dimensional ferrotoidic chains \cite{zhang2021,zhang2022}. Recently, we have also studied the magnetism in the ferromagnet La$_3$MnAs$_5$, and found that this compound consists of spin chains coupled via itinerant electrons \cite{duan2022}. Here we report the theoretical studies from first principles for a newly synthesized spin-chain compound Ba$_6$Cr$_2$S$_{10}$ (BCS) under high-pressure and high-temperature conditions. BCS has hexagonal lattice structure in which the spin chain consisting of Cr atoms is along the $c$-axis, as shown in Fig.\ref{fig:1}(a). In terms of chain structures, BCS has face-sharing octahedrons formed by sulfur atoms, which is similar to the AMX$_3$ and B$_3$MM'O$_6$. Our calculations show that the dimerisation along the chain induced mainly by the octahedrons formed by the S atoms will lead to anti-ferromagnetic ground state and thus ferro-toroidic ground state. In addition, our calculations show an both charge density wave and orbital ordering on Cr atoms along the chain due to the dimerized peculiar face-sharing octahedral structure. The small size of the Hubbard-$U$ parameter suggests a significant screening effect from the surrounding atoms. The remaining discussion falls into three sections. In \S\ref{sec:methods}, we introduce the computational methods. In \S\ref{sec:results}, we present our results and discuss them. In \S\ref{sec:conclusions}, we draw some general conclusions.
%- review 1d magnetic materials and toroidal moments.
%- review spin chain compounds.
%- review the previous work on similar compounds?
%- briefly introduce the work here.
%XXFIG.1: structure

%XXFIG.4: spin densities, orbital ordering

\section{Methods}\label{sec:methods}
First-principles density functional theory (DFT) calculations have been performed to study the electronic structure and magnetic properties of BCS using the projector augmented wave (PAW) method~\cite{paw} implemented in the VASP package~\cite{vasp1,vasp2}. The standard generalised-gradient approximation for the exchange-correlation functional - Perdew-Burke-Ernzerhof (PBE) - has been chosen for all the calculations \cite{pbe}. The kinetic energy cutoff of the plane-wave basis was set to be 350 eV. A $8\times8\times6$ $k$-point mesh was used for the Brillouin zone (BZ) sampling. The Fermi level was broadened by the Gaussian smearing method with a width of 0.05 eV. The self-consisten-field (SCF) converging process is accelerated by using Broyden's second method \cite{johnson1988}. For the structure without dimerisation, we have adopted the crystal structure of La$_3$MnAs$_5$ by replacing La, Mn, As using Ba, Cr, and S, and then optimized the crystal structure \cite{duan2022}. The band structure was plotted as the high-symmetry points in the first Brilloin zone of a hexagonal lattice, i.e., $\Gamma (0, 0, 0) \rightarrow M (\frac12, 0, 0)\rightarrow K (\frac23, \frac13, 0)\rightarrow \Gamma (0, 0, 0)\rightarrow A (0, 0, \frac12)\rightarrow L (\frac12, 0, \frac12) \rightarrow H (\frac23, \frac13, \frac12)\rightarrow A (0, 0, \frac{1}{2})$.

The symmetry group for BCS is $P\overline{6}2c$. The experimental lattice parameters are $a = b = 9.1371 \mathrm{\AA}$, $c = 12.3175 \mathrm{\AA}$, $\alpha = \beta = 90^\circ$, and $\gamma = 120^\circ$. In our calculations, the lattice constants and internal atomic positions were allowed to relax until all the forces on atoms were smaller than 0.01 eV/Å. The optimized lattice constants are: $a_o = b_o = 9.27 \mathrm{\AA}$ and $c_o = 12.44 \mathrm{\AA}$, which agree well with the experimental measurements~\cite{zhang2022}. Notice that there are two types of S atoms including (i) ligands for Cr (S1) and (ii) loosely bonded with Ba forming another chain (S2), as shown in Fig.\ref{fig:1}(a). We have also found the S atoms loosely bonded with Ba are non-uniformly dimerized, as shown in Fig.\ref{fig:1}(a). We have computed the intra-chain nearest-neighbouring (NN) and next-nearest-neighbouring (NNN) exchange interactions. Notice that the NN exchange interactions are different for the neighbouring exchange interactions; we distinguished them by $J_{\text{NN}}^1$ and $J_{\text{NN}}^2$ as labeled in Fig.\ref{fig:1}(a). The $U$-parameter has also been added (DFT+U) to work out the dependence of exchange interactions on $U$. A set of values of $U$ from $0$ to $2$ eV with increments of 0.5 eV were used for the DFT+U calculations. The spin-orbit interaction (SOI) was included into the spin-polarized calculations to understand the non-collinear magnetic configuration. In all our calculations we have kept the spin in the $ab$ plane. To validate our DFT+U+SOI results, we have also performed hybrid-exchange DFT calculations using HSE06 functional, implemented in VASP \cite{hse06}.

%- crystal structure
%- pseudo potentials
%- k-mesh
%- DFT + U
%- DFT + U + SOI

\section{Results and discussion}\label{sec:results}
%- band structure and density of states
%- exchange interactions.
%- spin densities and orbital ordering.
%- spin anisotropy and geometry dimerisation
%- geometry dimerisation optimisation.
%- etc

%% spin density and charge density figure

%% density of states for U = 0.5 eV

%% band gaps?

\begin{figure}[htbp]
%\begin{tabular}{cc}
%\includegraphics[width=9cm,height=5cm]{fig_1.eps}\\
\includegraphics[width=9cm, height=4.0cm, trim={0cm 10.5cm 0.0cm 0.5cm},clip]{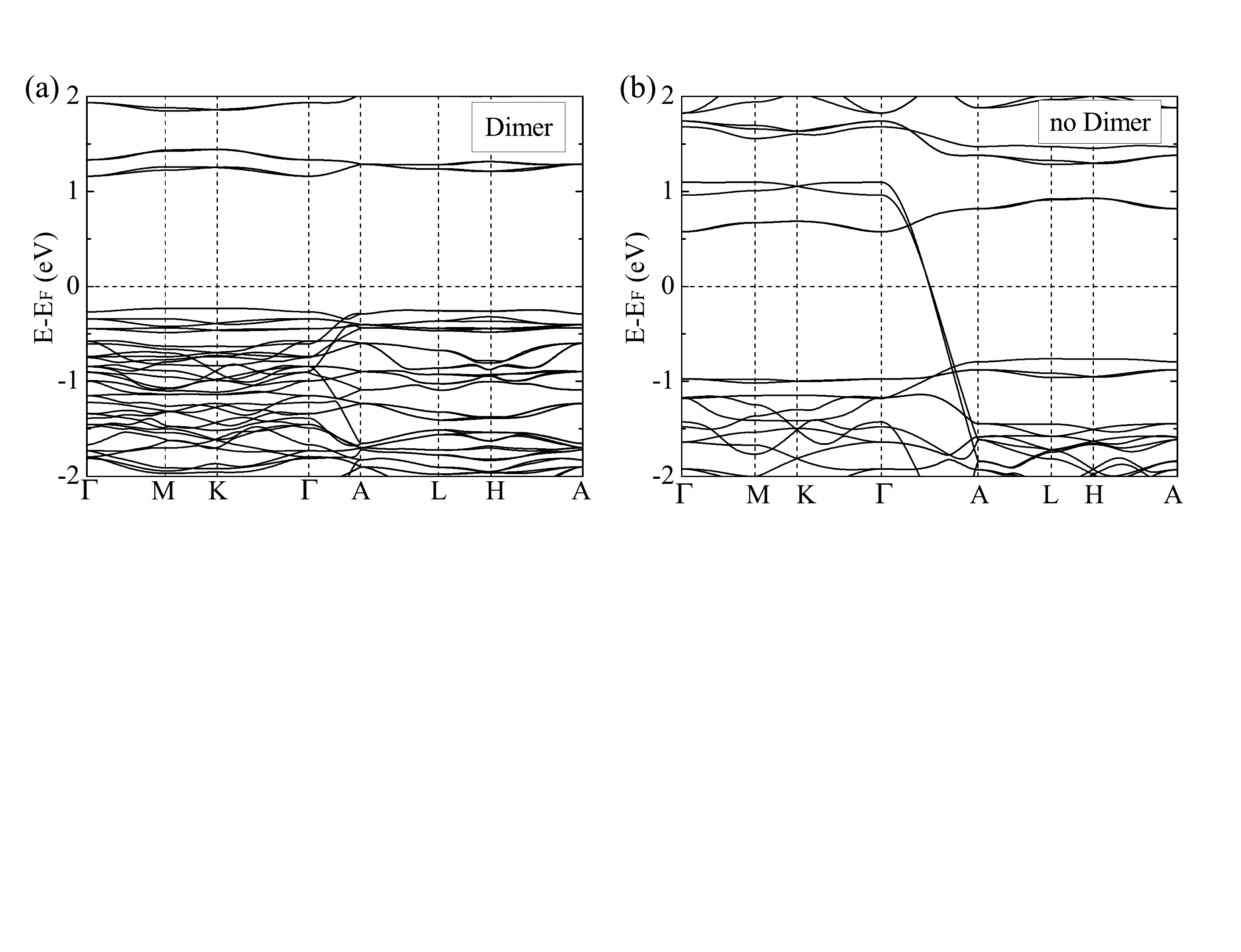}\\
%\textbf{(a)}\\
%\includegraphics[width=8cm,height=7cm]{}\\
%\includegraphics[width=18cm, height=12.35cm, trim={0.cm 0cm 0cm 0cm},clip]{fig_1b.pdf}\\
%\textbf{(b)}\\
%\end{tabular}
\caption{(Colour online.) A comparison between the band structures of the crystal structures with dimerisation and without. We can clearly see the clear Peierls transition to a CDW, which is due to the S ligands rather than Cr atoms. The calculations here were performed without $U$.}\label{fig:2}
\end{figure}

The computed band structures for BCS with dimerisation and without are shown in Fig.\ref{fig:2} (a) and (b), respectively. The optimized dimerisation for Cr was only $0.005 \ \mathrm{\AA}$ in our calculations. By contrast, the dimerized S2-type atoms play a dominant role for a magnetic insulating state. In Fig.\ref{fig:2}(b), for the BCS structure without dimerisation, we can see a significant band dispersion originating from sulfur atoms crossing the band gap, leading to a metallic state. The band dispersion across the band gap is along $\Gamma(0, 0, 0)\rightarrow A (0, 0, \frac12)$ - the chain direction, which is consistent with the dimerization that is also along the $c-$axis. The dimerisation of S2 atoms will open up a band gap - a typical characteristics for the Peierls transition to a charge density wave (CDW) phase in one dimension \cite{fro1954,gruner1988}, which is closely related to quantum fluid state in 1D. This CDW phase has been illustrated clearly by the atomic charge labeled in Fig.\ref{fig:1}(a). The atomic charges for the S2 atoms are 3.74$e$, 3.74$e$, 3.66$e$, and 3.66$e$, respectively. This unequal charges are due to the nonuniform dimerizations; the distance between the top (bottom) two S atoms is 2.61 (2.23) $\AA$. The dimerization also brings forward localized spins that can be seen clearly in Fig.\ref{fig:1}(b), suggesting the orbital ordering on Cr along the chain.

%Therein, not only charge ordering but also orbital ordering on Cr atoms can be formed due to the zig-zag lattice formation along the chain. Moreover, this CDW phase co-exists with the non-collinear magnetic properties in BCS, which will be discussed later on.

%XXFIG.2: band structure

\begin{figure}[htbp]
%\begin{tabular}{cc}
%\includegraphics[width=9cm,height=5cm]{fig_3.eps}\\
\includegraphics[width=8cm, height=6cm, trim={0cm 1.5cm 0.0cm 0.0cm},clip]{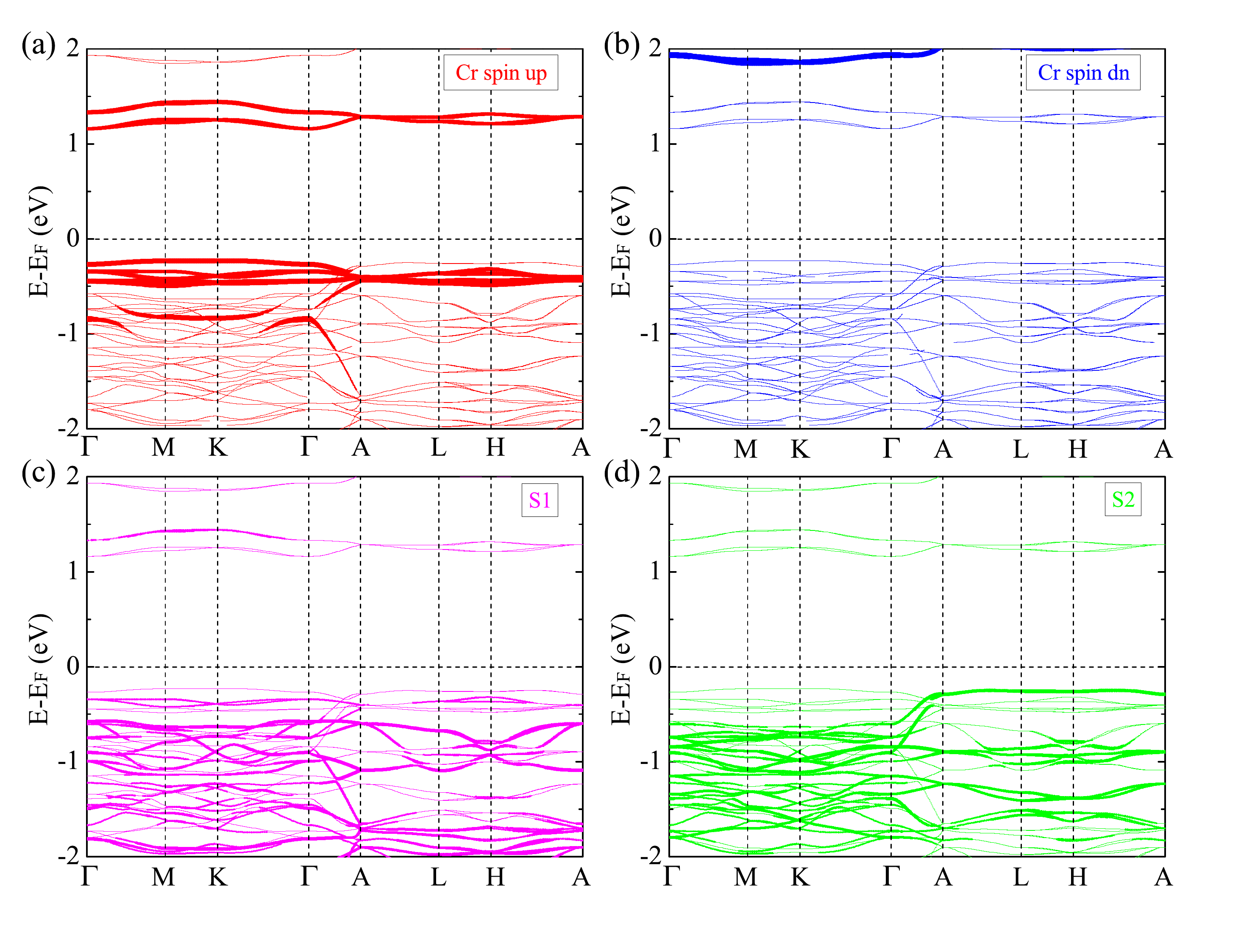}\\
%\textbf{(a)}\\
%\includegraphics[width=8cm,height=7cm]{}\\
%\includegraphics[width=18cm, height=12.35cm, trim={0.cm 0cm 0cm 0cm},clip]{fig_1b.pdf}\\
%\textbf{(b)}\\
%\end{tabular}
\caption{(Colour online.) The spin-polarised band structures projected to Cr and S, which clearly dominates the bands near the Fermi energy. }\label{fig:3}
\end{figure}

We have also computed the spin-polarised projected band structure for BCS with dimerisation, as shown in Fig.\ref{fig:3}. The spin-up band structures are plotted in (a), while the spin-down band structures are plotted in (b). For (c) and (d), we have contributions from both spin-up and spin-down for S1 and S2 projections, respectively. The band structure projected onto the Cr atoms for spin up is shown in (a), in which the dominant projections are at $E-E_F \sim -0.5$ eV, i.e., 0.5 eV below the Fermi surface. The projections to the S1 and S2 types of atoms (Fig.\ref{fig:3}c and d) labelled in Fig.\ref{fig:1}(a) suggest that S1 and S2 atoms contribute to the same energy range as Cr atoms, thus leading to strong hybridisation between Cr and S1-type atoms. 

%XXFIG.3: projected band structure

We found that the magnetic anisotropy energies (MAE) calculated here as the total-energy difference between the spin configurations along $c$- and $a$-axes. The energy difference computed here in VASP is comparable to 0.1 meV, which is beyond the limit of the general accuracy for the DFT calculations by using VASP. We therefore leave this for further investigation. We speculate that the high accuracy of VASP could lead to a small energy difference. In addition, this nearly degenerate magnetic ground state ($< 1$ meV) could lead to interesting physics through quantum fluctuations between these two spin configurations.

\begin{table}[t]
\caption{Exchange interactions,band gaps and $d$-orbital levels near the Fermi energy as a function of Hubbard-$U$ parameters. }
\begin{center}
\begin{tabular*}{8cm}{@{\extracolsep{\fill}} cccccc}
\hline\hline
$U$ (eV) & 0 & \textcolor{blue}{0.5} & 1.0 & 1.5 & 2.0\\
\hline
% Band gap (eV) &1.39&1.56&1.65&1.67&1.68\\
% \hline
$J_\mathrm{NN}^1$ (meV) & 14.5 & \textcolor{blue}{11.7} & 9.4 & 7.5 & 5.9\\
$J_\mathrm{NN}^2$ (meV) & 17.0 & \textcolor{blue}{14.0} & 11.5 & 9.3 & 7.6\\
$J_\mathrm{NNN}$ (meV) & 0.7 & \textcolor{blue}{0.5} & 0.2 & 0.1 & 0 \\
\hline
Band gap (eV) &1.00&\textcolor{blue}{1.04}&1.06&1.08&1.09 \\
\hline
$d$-orbital lower levels (eV) &-0.4&\textcolor{blue}{-0.6}&-0.7&-0.9&-1.1 \\
$d$-orbital upper levels (eV) &1.3&\textcolor{blue}{1.3}&1.3&1.3&1.3 \\
$d$-orbital gap (eV) &1.7&\textcolor{blue}{1.9}&2.0&2.2&2.4 \\
\hline\hline
\end{tabular*}\label{tab:1}
\end{center}
\end{table}

We have also performed the DFT+U+SOI calculations implemented in VASP to study the dependence of exchange interactions between the spins on the Cr ions on the on-site Coulomb $U$ parameter. Here we have adopted a spin Hamiltonian as follows, 

\begin{eqnarray}\label{eq:1}
\hat{H} &=& \sum_{\langle i,j\rangle\in \mathrm{NN,1}}J^1_{\mathrm{ NN}}\hat{s}_i\cdot\hat{s}_j + \sum_{\langle i,j\rangle\in \mathrm{NN,2}}J^2_{\mathrm{NN}}\hat{s}_i\cdot\hat{s}_j\\\nonumber
&&+\sum_{\langle i,j\rangle\in \mathrm{NNN}}J_{\mathrm{NNN}}\hat{s}_i\cdot\hat{s}_j    
\end{eqnarray}
Positive exchange interaction implies AFM coupling, where negative for FM couplings. To extract the exchange interactions in eq.\ref{eq:1}, we have performed the total energy calculations for four spin configurations on Cr atoms, including $\ket{1}=\ket{\uparrow,\uparrow,\uparrow,\uparrow}$, $\ket{2}=\ket{\uparrow,\downarrow,\uparrow,\downarrow}$, $\ket{3}=\ket{\uparrow,\downarrow,\downarrow,\uparrow}$, and $\ket{4}=\ket{\uparrow,\uparrow,\downarrow,\uparrow}$. Our calculations confirm that the state $\ket{2}$ is the ground state, which is consistent with the experimentally determined magnetic structure and calculations for BCS \cite{zhang2022}. The exchange interactions in eq.\ref{eq:1} can then be computed according to
\begin{eqnarray}
    J_{\text{NN}}^1 && = -\frac{\Delta E_{31}}{4 S^2}-2J_{\text{NNN}}\\\nonumber
    J_{\text{NN}}^2 && = -\frac{\Delta E_{21}}{4 S^2}-J_{\text{NN}}^1 \\\nonumber
    J_{\text{NNN}} && = \frac{\Delta E_{21}}{8 S^2}-\frac{\Delta E_{41}}{4 S^2}
\end{eqnarray}

%XXFIG.5: hubbard u
\begin{figure}[htbp]
%\begin{tabular}{cc}
%\includegraphics[width=9cm,height=5cm]{fig_3.eps}\\
\includegraphics[width=8cm, height=5.2cm, trim={0cm 0cm 0.0cm 0.0cm},clip]{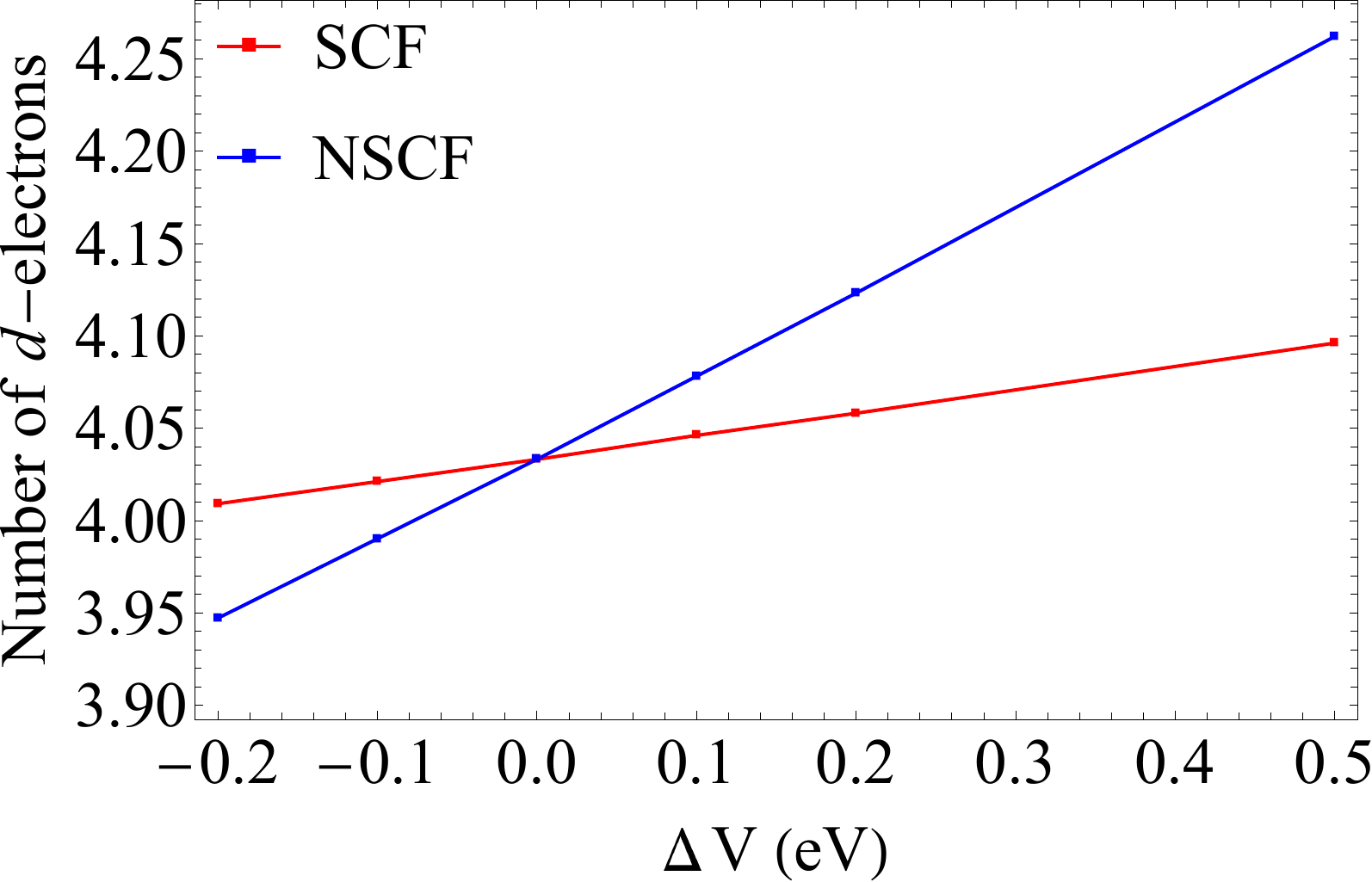}\\
%\textbf{(a)}\\
%\includegraphics[width=8cm,height=7cm]{}\\
%\includegraphics[width=18cm, height=12.35cm, trim={0.cm 0cm 0cm 0cm},clip]{fig_1b.pdf}\\
%\textbf{(b)}\\
%\end{tabular}
\caption{(Colour online.) We have computed the bare on-site Coulomb interaction $U_{\text{bare}}$ by using the linear response of the electron populations on $d$-orbitals as shown in the figure. The SCF (NSCF) calculations are plotted in red (blue). The gradients of the lines correspond to the charge susceptibilities.}\label{fig:4}. 
\end{figure}

%XXFIG.4: dos
\begin{figure}[htbp]
%\begin{tabular}{cc}
%\includegraphics[width=9cm,height=5cm]{fig_3.eps}\\
\includegraphics[width=8cm, height=5.5cm, trim={0cm 0cm 0.0cm 0.0cm},clip]{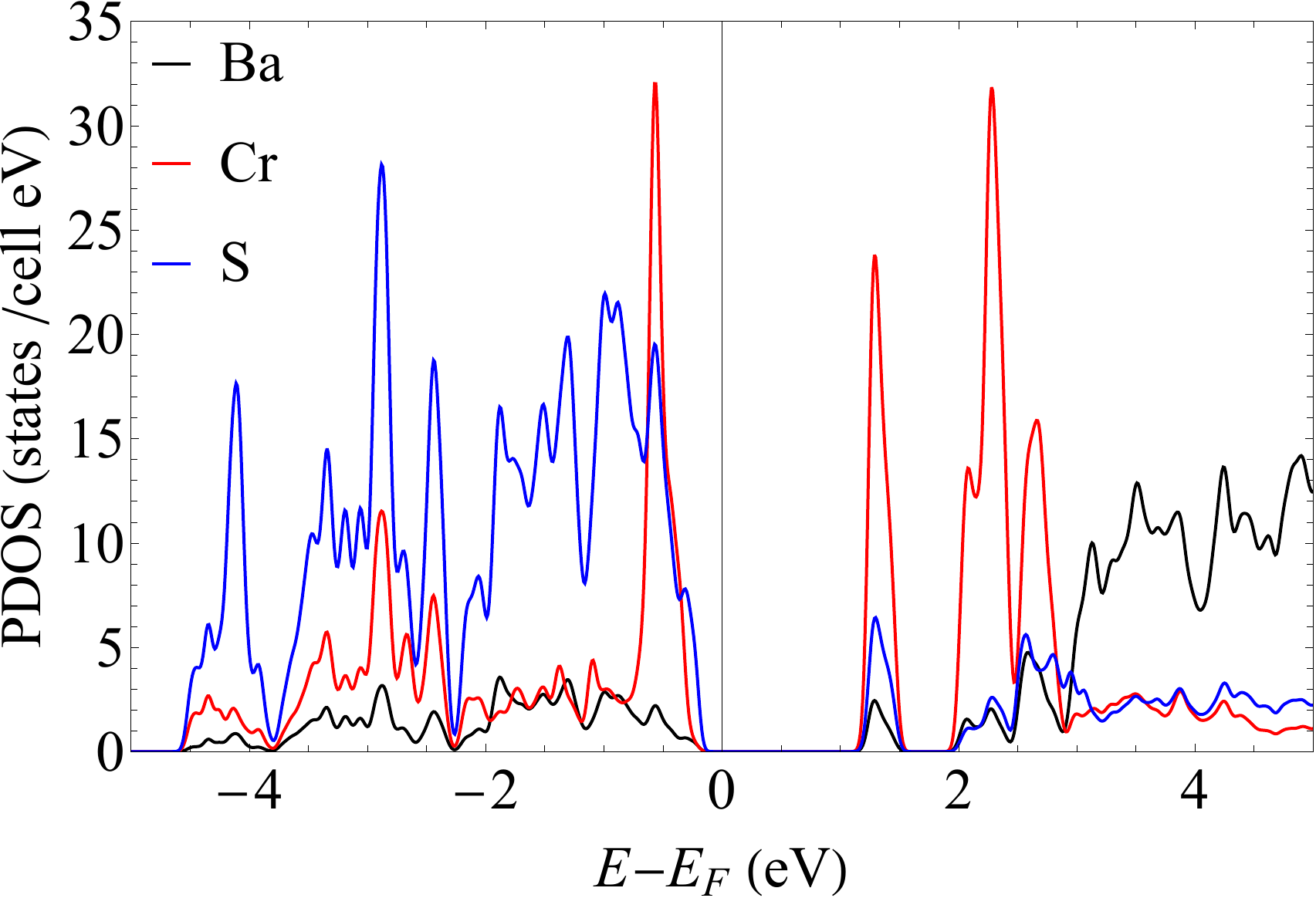}\\
%\textbf{(a)}\\
%\includegraphics[width=8cm,height=7cm]{}\\
%\includegraphics[width=18cm, height=12.35cm, trim={0.cm 0cm 0cm 0cm},clip]{fig_1b.pdf}\\
%\textbf{(b)}\\
%\end{tabular}
\caption{(Colour online.) The PDOS projected onto Ba (black), Cr (red), and S (blue) is shown. We can see the significant overlaps between Cr and S near the Fermi energy, suggesting strong bondings between them. A further detailed check at the Fermi energy suggests S is dominant.}\label{fig:5}. 
\end{figure}

All the computed band gaps ($\sim 1$ eV) are larger than the experimental observed ($0.559$ - $0.843$ eV) as shown in Table.\ref{tab:1}. This discrepancy could be due to a few aspects. The functional used could overestimate the dimerization during the geometry optimization process. The accuracy of the experimental band gap could be affected by defects. We can also see the very weak dependence of the band gap on the $U$-parameter, suggesting the band gap is dominated by dimerized S1 atoms and associated electron hopping strength. Moreover, based on the comparison of band gap, we could not find the optimal $U$ parameter through band-gap calculations. Moreover, we have also computed the projected density of states (PDOS) for all the atomic species, as shown in Fig.\ref{fig:4}. We have found that the contributions from S atoms are dominant at the Fermi energy. The weak dependence of band gap on $U$ togehther with the PDOS suggests sulfur contributions are dominant over the Fermi energy and the band gap, which is consistent with the picture in which sulfur dimerization induces CDW.

The calculated exchange interactions were tabulated in Table.\ref{tab:1}. We can see that the intra-chain exchange interactions will decreases as $U$ increases, which implies the exchange mechanism could be dominated by super-exchange ($\propto\frac{1}{U}$). The NN and NNN exchange interactions along the chain are AFM, which suggests a weak spin frustration in one dimension. Due to the strong quantum fluctuations in 1D, this could result in quantum spin liquid. On the other hand, we can also derive the experimental intra-chain coupling from the magnetic measurement of $\theta_p$ = -714 K \cite{zhang2022} by using the mean-field expression for $\theta_p =- \frac{2S(S+1)}{3}(J_{\text{NN}}^1+J_{\text{NN}}^2-2J_{\text{NNN}})$ \cite{nrbook, schron2010} with $S = \frac{3}{2}$. Based on this mean-field approach, we have found the optimal effective Hubbard-$U$ parameter ($U_{\text{opt}}$) is $\sim 0.5$ eV. In the mean time, we have also computed the bare $U$ parameter ($U_{\text{bare}}$) by using the linear-response method described in Ref.\cite{coco2005}, i.e., $U_{\text{bare}} = \frac{1}{\chi}-\frac{1}{\chi_0}$. Here $\chi = \frac{dQ}{dV}$ is the SCF charge response to the change of electric spherical potential applied to the $d-$orbitals. On the other hand, $\chi_0$ is the NSCF charge response. The gradients of the red ($\chi$) and blue ( $\chi_0$) lines in Fig.\ref{fig:4} are $0.124$ and $\sim 0.451$, respectively, leading to a $U_{\text{bare}} = 5.8$ eV. We can see that $U_{\text{opt}}$ is one order smaller than $U_{\text{bare}}$, which could be due to the screening effect from the Coulomb interaction between sites ($V$) and Hund's exchange interaction $J_\text{H}$. The recent estimate suggested $J$ for Cr is between $0.5$ and $0.93$ eV \cite{ms2020,gl2014}. An estimate of $V$ with a Cr-Cr distance $\sim 3.1 \ \AA$, which will lead to a Coulomb interaction $\sim \frac{1}{6}\text{Hartree} \simeq 4.5$ eV. Therefore, $U_{\text{opt}} \simeq U_{\text{eff}} = U_{\text{bare}}-V-J_\text{H}$, in which the effective Hubbard $U$ results from the screening effect from inter-site Coulomb interaction and Hund's exchange. Our further calculations with $U = 5.8$ eV show that the ground-state magnetic structure will change to the ferromagnetic state ($\ket{1}$), which is inconsistent with the experimental observation \cite{zhang2022}. Another theoretical work based on basis-independent constrained random-phase approximation \cite{akj2006} has suggested the fully screened $U$-parameter for Cr due to metallic elements is $\sim 1$ eV for Y(Sr)XO$_3$ (X = transition metals), further supporting the small size of $U_{\text{eff}}$ here if taking into account the Hund's exchange (in the order of 0.1 eV). In addition, we have summarized the $d$-orbital levels in Table.\ref{tab:1} and the gap between highest occupied (lower) and the lowest unoccupied (upper) $d$-bands. This gap is also rather weakly dependent on the $U$ parameter, suggesting the existence of screening effect on $U$. This screening effect reduces the effective on-site Coulomb interaction, strengthening the super-exchange interaction ($\propto\frac{1}{U_{\text{eff}}}$), thus leading to an AFM ground state. The computed bandwidth (W) for the upper $d$-band is $\simeq0.4$ eV, resulting in a small ratio of $\frac{U}{W} \simeq 1.25$.  Thus, owing to the small size of $U$, BCS could be close to the boundary between correlated metal and Mott insulator, in which there is rich physics \cite{dmft} such as quantum spin liquid. Our HSE06 DFT calculations further confirmed our DFT+U+SOI results from the following three perspectives. First, the exchange interactions computed by HSE06 is AFT with a magnitude of $33$ meV, consistent with the DFT+U+SOI calculation. The overestimation of the exchange interaction is characteristic for hybrid-exchange functional \cite{stroppa2008}. Secondly, based on the analysis of forces on the S atoms along the S2 chain, we have found similar geometry formation of the chain, thus also leading to a CDW along S2, which supports our DFT+U+SOI calculations. Thirdly, the band gap computed using HSE06 is $\sim 0.6$ eV, which agrees well with the measurement shown in the Appendix.
 
\ 
\ 

\section{Conclusions}\label{sec:conclusions}
In summary, we have computed the electronic structure and magnetic properties of the spin chain compound Ba$_6$Cr$_2$S$_{10}$ based on DFT+U+SOI methods. Our calculations suggest an electrostatic screening effect reduces $U$ significantly ($U \simeq 0.5$ eV), leading to an AFM ground state. The spin densities on Cr atoms suggest an orbital ordering along the Cr chain. In addition, the sulfur atoms loosely bonded with Ba play an important role in the dimerisation, which leads to a CDW phase and an insulating ground state. Ba$_6$Cr$_2$S$_{10}$ could be close to the boundary between the Mott insulator and correlated metal due to the small effective $U$. Overall we found the co-existence of CDW and AFM along different chains, which is not only rare in the literature, but also provides independent tunability over charge and spin degrees of freedom. In addition, the AFM NNN exchange interaction could lead to spin frustration and hence quantum spin liquid state.

\section*{Data Availability}
All the data that support the findings of this study are available from the corresponding author upon reasonable request.

%There are a few aspects which can be improved in the further studies. The first is the role of quantum fluctuations to the Curie intercept. Previously the cobalt-phthalocyanine spin chains have been studied by using finite-size Heisenberg spin chain model \cite{serri2014}, in which the experimental exchange interactions are consistent with the first principles calculations \cite{wu2013}. Therefore, the simulation of the Curie intercept based on a quantum spin-chain model might be needed to fit a more accurate exchange interaction, at least for the intra-chain NN ones. Moreover, the electron-phonon interaction is required for the formation of CDW, which can be investigated in the future \cite{gruner1988}.

%\newpage

\section*{ACKNOWLEDGEMENTS}
JHZ, YLZ, and JC acknowledge the funding from the National Natural Science Foundation of China under Grant No. 92165101.
WW wishes to acknowledge the funding support from STFC UK. 
\newline

\section*{Author Contributions}

WW contributed to the concept of the paper. JHZ, JFZ, WW, YLZ, and JC performed the theoretical analysis. DA, XCW and CQJ contributed to the experimental part. All the authors wrote the paper. 

\section*{Competing Interests}
The authors declare no competing interests.

%We thank the inspiring discussion with Prof. Changqing Jin and Prof. Xiancheng Wang.
%We thank Prof. Xiancheng Wang and Prof. Changqing Jin for helpful and inspiring discussions.
\newpage

%\begin{appendix}
%\setcounter{figure}{0} 
\appendix
\counterwithin{figure}{section}

\section{Band gap measurement}
The band gap of BCS has been determined experimentally by measuring the electrical resistance as a function of temperature. We have measured the resistance at rather high temperature (375-400 K) as it is too large to measure at lower temperature. Then we fitted the data to a model $\ln(R)=\ln(R_0) + E_g/2k_BT$, where $R$ is the resistance, $E_g$ is the band gap, $k_B$ is the Boltzmann constant, and $T$ is temperature. We have fitted the first and second half of data using two separate linear model, which shows the band gap of BCS is between $0.559\pm 0.009$ and $0.843\pm 0.004$ eV.
%XXFIG.6: band gap
\begin{figure}[htbp]
%\begin{tabular}{cc}
%\includegraphics[width=9cm,height=5cm]{fig_3.eps}\\
\includegraphics[width=7cm, height=5.cm, trim={0cm 0cm 0.0cm 0.0cm},clip]{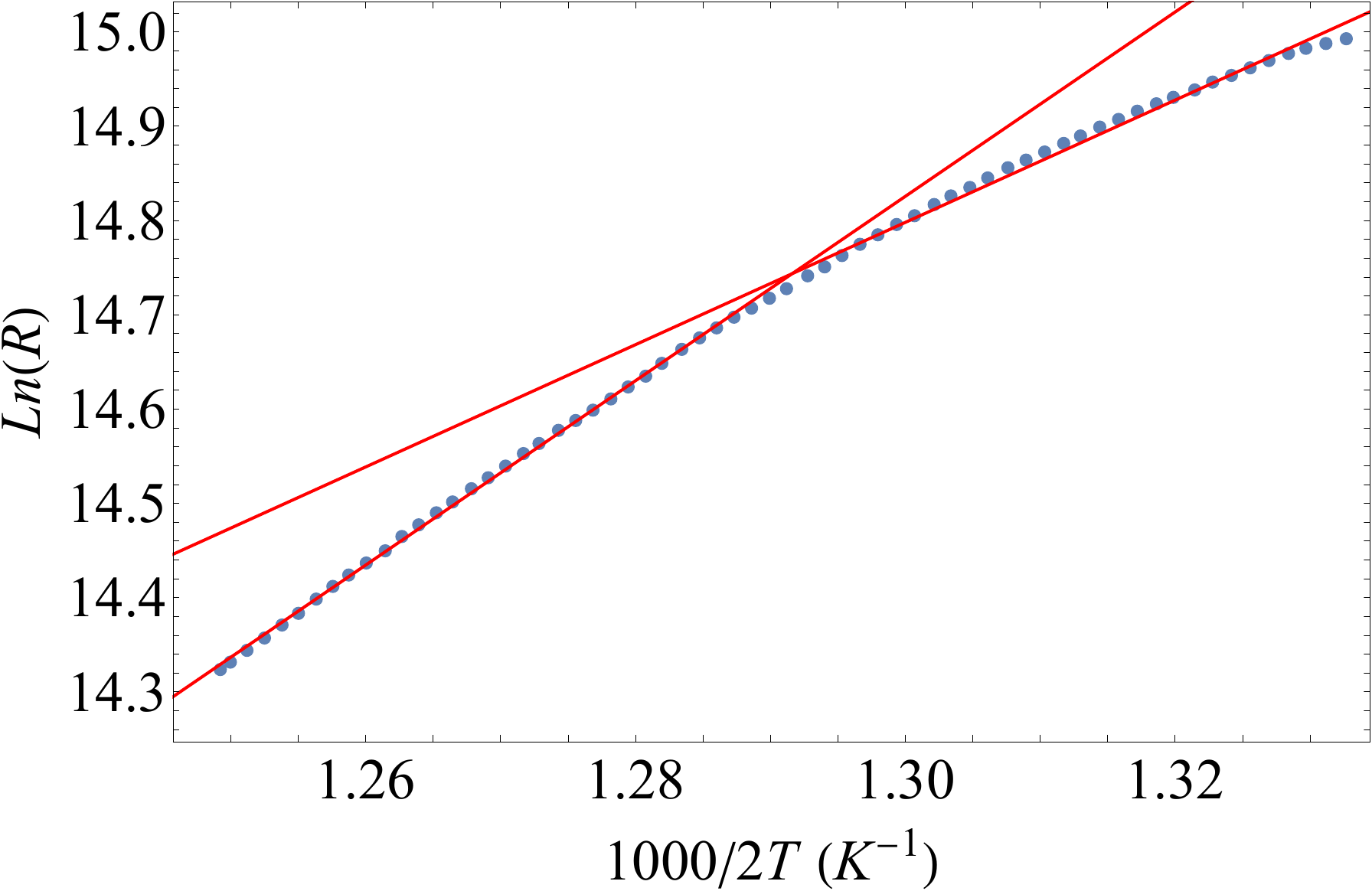}\\
%\textbf{(a)}\\
%\includegraphics[width=8cm,height=7cm]{}\\
%\includegraphics[width=18cm, height=12.35cm, trim={0.cm 0cm 0cm 0cm},clip]{fig_1b.pdf}\\
%\textbf{(b)}\\
%\end{tabular}
\caption{(Colour online.) The experimental data (blue points) are linearly fitted separately. The two gradients of the fittings are $6.485\pm 0.108$ and $9.777\pm 0.051$, leading to a band-gap range between $0.559\pm 0.009$ and $0.843\pm 0.004$ eV.}\label{fig:5}
\end{figure}

%\end{appendix}

\end{document}